\documentclass[letter,twocolumn]{jpsj3}
\usepackage{graphicx}
\usepackage{amsmath}
\usepackage{braket}
\usepackage{bm}
\usepackage{color}

\title{Entanglement Chern Number of the Kane--Mele Model with Ferromagnetism}

\author{Hiromu Araki$^1$, Toshikaze Kariyado$^{1,2}$, Takahiro Fukui$^3$, and Yasuhiro Hatsugai$^{1,2}$}
\inst{$^1$Graduate School of Pure and Applied Science, University of
Tsukuba, Tsukuba, Ibaraki 305-8571, Japan\\
$^2$Division of Physics, University of
Tsukuba, Tsukuba, Ibaraki 305-8571, Japan\\
$^3$Department of Physics, Ibaraki University, Mito, Ibaraki 310-8512, Japan\\
}

\abst{The entanglement Chern number, the Chern number for the
entanglement Hamiltonian, is used to characterize the Kane--Mele model,
which is a typical model of the quantum spin Hall phase with 
time-reversal symmetry. We first obtain the global phase diagram of the
Kane--Mele model in terms of the entanglement spin Chern number, which is
defined by using a spin subspace as a subspace to be traced out in
preparing the entanglement Hamiltonian. We
further demonstrate the effectiveness of the entanglement Chern
number without time-reversal symmetry
by extending the Kane--Mele model to include the Zeeman
term. The numerical results confirm that the sum
of the entanglement spin Chern number is equal to the Chern number. 
}

\begin{document}
\maketitle

%\section{Introduction}

Symmetry enriches topological classification of material phases.\cite{PhysRevB.76.045302}
For
free fermion systems, the fundamental symmetries, i.e., 
time-reversal symmetry, charge conjugation symmetry, 
and chiral symmetry, are essential to
obtain the so-called periodic table of topological insulators and
superconductors.\cite{PhysRevB.78.195125,PhysRevB.78.195424,DOI:10.1063/1.3149495,1367-2630-12-6-065010}
The classification has been farther refined to
include some crystalline point group symmetries.\cite{PhysRevLett.106.106802,
PhysRevB.88.125129,PhysRevB.90.165114} In some cases, the physical
and intuitive construction of topologically nontrivial phases with
higher symmetry is possible by assembling two or multiple copies of
topologically nontrivial phases with lower symmetry so that the symmetry
of the assembled system is restored. A typical example is a quantum
spin Hall (QSH) insulator with time-reversal
symmetry.\cite{PhysRevLett.95.146802,PhysRevLett.95.226801}
Physically, it is constructed by making each spin subsystem
(up and down) a quantum Hall state.\cite{PhysRevLett.97.036808,PhysRevB.75.121403}
The point
is that, even when the whole system has time-reversal symmetry, the
symmetry is effectively broken and the Chern number is finite for
each spin subspace.

The ``entanglement'' Chern number has recently been 
introduced to characterize
various topological ground states \cite{doi:10.7566/JPSJ.83.113705}. The
entanglement Chern number is the Chern number\cite{PhysRevLett.49.405,
PhysRevLett.61.2015,PhysRevLett.71.3697}
for the entanglement
Hamiltonian, and the entanglement Hamiltonian is constructed by tracing
out certain subspaces of a given system.\cite{PhysRevB.84.195103,PhysRevLett.113.106801,
doi:10.7566/JPSJ.84.043703}
This means that the entanglement
Chern number is suitable for analyzing the topological properties of a
high-symmetry system composed of multiple copies of lower-symmetry
systems. That is, we can focus on a specific subsystem by tracing out
the others. For instance, if the up- or down- spin sector is
chosen as a subspace to be traced out, the obtained entanglement Chern
number, which we name as \textit{the entanglement spin Chern number},
should be useful for characterizing the QSH state. It is worth noting that
the choice of the subspace is not limited to spin sectors and that
entanglement Chern number potentially has wide applications. Also, the
entanglement Chern number can be defined regardless of the symmetry of
the system or the details of the Hamiltonian provided we can choose a
subsystem to be traced out. 

In this paper, we first briefly explain the idea behind the
entanglement (spin) Chern number. Then, we extend the arguments
in Ref.~\citen{doi:10.7566/JPSJ.83.113705} to cover the global phase
diagram of the Kane--Mele model, which is a typical model for the QSH
state.  We also investigate the stability of the entanglement spin Chern
number against time-reversal symmetry breaking by introducing the
Zeeman term to the Kane--Mele model. 
It is found that the sum of the entanglement spin Chern numbers 
is equal to the Chern number in the entire phase space.
In addition, in the strong spin-orbit
coupling limit, a phase with a large (up to three) magnitude of the
Chern number is shown to appear.

%\section{Entanglement Chern number}
Let us first introduce the entanglement Chern number, which plays a key
role in this paper. Briefly speaking, the entanglement Chern number is
the Chern number for the entanglement Hamiltonian. In order to define the
entanglement Hamiltonian, we divide a given system into two
subsystems, say $A$ and $B$. Then, the entanglement Hamiltonian for this
partition, $H_A$, is defined as
$\mathrm{e}^{-H_A}=\rho_A\equiv\mathrm{Tr}_B\rho$
with $\rho\equiv|\Psi\rangle\langle\Psi|$, where $|\Psi\rangle$ and
$\mathrm{Tr}_B$ denote the ground-state wave function and the trace over
subsystem $B$, respectively. The name ``entanglement'' Hamiltonian
originates from the fact that information of the entanglement 
between $A$ and
$B$ is encoded in $\rho_A$\cite{PhysRevB.73.245115} or equivalently in $H_A$. 

In general, a given Hamiltonian
$H = \sum_{i,j} c_i^\dagger\mathcal{H}_{ij}c_j$ and a correlation matrix
$C_{ij}\equiv\langle{c_i^\dagger c_j}\rangle=\mathrm{Tr}[\rho
c_i^\dagger c_j]$ are related as\cite{0305-4470-36-14-101}
\begin{equation}
 \label{eq:HtoC}
 \mathcal{H}^T = \mathrm{ln}[(1-C)/C].
\end{equation}
Moreover, at zero temperature, the correlation matrix is
explicitly written as
\begin{equation}
 \label{eq:Corre}
 C_{ij} = \sum_{n:\text{ occupied}}\phi_n^*(i)\phi_n(j),
\end{equation}
where $\phi_n(i)$ is the eigenvector of $\mathcal{H}$. Now, we define
the restricted correlation matrix $C_A$ by projecting $C$ to subsystem
$A$. Namely, the elements of $C_A$ are taken from $C_{ij}$ with $i$ and
$j$ in $A$. As in the general case of Eq.~\eqref{eq:HtoC}, we have
$\mathcal{H}_A^T = \ln[(1-C_A)/C_A]$, which gives a convenient way
to evaluate $\mathcal{H}_A$. 
In the following, we use the spectrum of $C_A$ to evaluate 
the spectrum of $\mathcal{H}_A$ and call it the
``entanglement spectrum''.

When the considered partition into $A$ and $B$ retains the
translation symmetry of the original model, the momentum $\bm{k}$ also
becomes a good quantum number for the entanglement Hamiltonian. 
Then, the matrix to be analyzed becomes
\begin{equation}
 C_A(\bm{k}) = P_A P_{-}(\bm{k})P_A,
 \label{eq:reduced-corre}
\end{equation}
where $P_{-}(\bm{k})=\sum_{n:\text{ occupied}}\tilde{\phi}^\dagger_n(\bm{k})
\tilde{\phi}_n(\bm{k})$ is the projection operator to the occupied bands
defined using the Bloch wave function
$\tilde{\phi}^\dagger_n(\bm{k})$
for $n$th band and $P_A$ is the projection operator to the subsystem $A$.
In this case, we obtain a momentum-resolved entanglement Hamiltonian
$\mathcal{H}_A(\bm{k})$ as
$\mathcal{H}_A^T(\bm{k})=\mathrm{ln}[(1-C_A(\bm{k}))/C_A(\bm{k})]$.
This relation means that $\psi_{C_A}^*(\bm{k})$ is
an eigenvector of
$\mathcal{H}_A$ if $\psi_{C_A}(\bm{k})$ is an eigenvector of
$C_A(\bm{k})$. Therefore, it is possible to define the (entanglement)
Chern number by using the eigenvector $\psi_n(\bm{k})$ of $C_A(\bm{k})$,
which is given as a solution of
$C_A(\bm{k})\psi_n(\bm{k})=\xi_n(\bm{k})\psi_n(\bm{k})$. The eigenvalues
$\xi_n(\bm{k})$ form a band structure, and the values of 
$\xi_n(\bm{k})$ are restricted in the range $[0,1]$. 
If there is a finite gap in the band structure of $\xi_n(\bm{k})$,
for example between the $l$th and $(l+1)$th bands, we can define the (nonabelian)
Berry connection
$A_\mu(\bm{k})=\psi^\dagger(\bm{k})\partial_\mu\psi(\bm{k})$ and the
curvature $F_{12}(\bm{k}) =
\partial_1A_2(\bm{k})-\partial_2A_1(\bm{k})$ for this gap. Here,
$\psi(\bm{k})$ is
the multiplet that consists of $\psi_n(\bm{k})$ for up to the 
$l$th band. Using these expressions, the entanglement Chern number is
defined as
\begin{equation}
 c_A = \frac{i}{2\pi}\int F_{12}(\bm{k})d^2k.
\end{equation}
Precise and efficient evaluation of the Chern number is made possible by
using link variables\cite{doi:10.1143/JPSJ.74.1674}.

The amount of information that can be extracted 
from the entanglement Chern number
crucially depends on the partition. One possible choice is the
partition into the spin-up and spin-down sectors. The entanglement
Chern number defined with such a partition is named the entanglement spin
Chern number, and it is considered to be 
useful for distinguishing the QSH insulator from ordinary
insulators\cite{doi:10.7566/JPSJ.83.113705,doi:10.7566/JPSJ.84.043703}.

%\section{Entanglement Chern number of Kane--Mele model}
The Kane--Mele model is a typical model for QSH
states\cite{PhysRevLett.95.226801}, whose
Hamiltonian is explicitly written as 
\begin{multline}
 \mathcal{H}_{KM} = t\sum_{\langle ij \rangle}{c_i^\dagger c_j}
  + i \lambda_{SO} \sum_{\langle \langle ij \rangle \rangle}
  {\nu_{ij} c^\dagger_i \hat{s}^z c_j} \\
  + i \lambda_{R} \sum_{\langle ij \rangle} c_i^\dagger
  \{ \bm{s} \times \bm{d}_{ij} \}^z c_j
  + \lambda_{\nu} \sum_{i} {\xi_i c_i^\dagger c_i}
\end{multline}
using $c_i={}^t\!\left(c_{i,\uparrow}~ c_{i,\downarrow}\right)$,
where $c_{i,\sigma}$
is the annihilation operator of a spin-$\sigma$ electron at the $i$th
site on the honeycomb lattice and $\bm{s}$ denotes the spin
operator. $\langle{ij}\rangle$ and $\langle\langle{ij}\rangle\rangle$
denote summation over the nearest-neighbor 
and the next-nearest-neighbor pairs of
sites, respectively. 
The first term is a nearest neighbor-hopping term on the honeycomb lattice.
The second term represents a spin-orbit coupling that is essential for
the QSH effect in this model, where $\nu_{ij}$ takes $\pm 1$ depending on
$i$ and $j$. The third and fourth terms are Rashba and
staggered potential terms, respectively. Here, $\hat{\bm{d}}$ is the
direction vector from the $i$th site to the $j$th site and
$\xi_i=\pm 1$. The Kane--Mele model is a four-band model where the four
degrees of freedom originate from two sublattices and two spins. When we
consider the entanglement spin Chern number, the entanglement Hamiltonian
gives two bands, since two degrees of freedom are traced out. Thus, the
entanglement spin Chern number is well defined if the 
entanglement bands are
nondegenerate within the whole Brillouin zone. 

\begin{figure}[htb]
 \centering
 \begin{center}
  \includegraphics[scale=0.4]{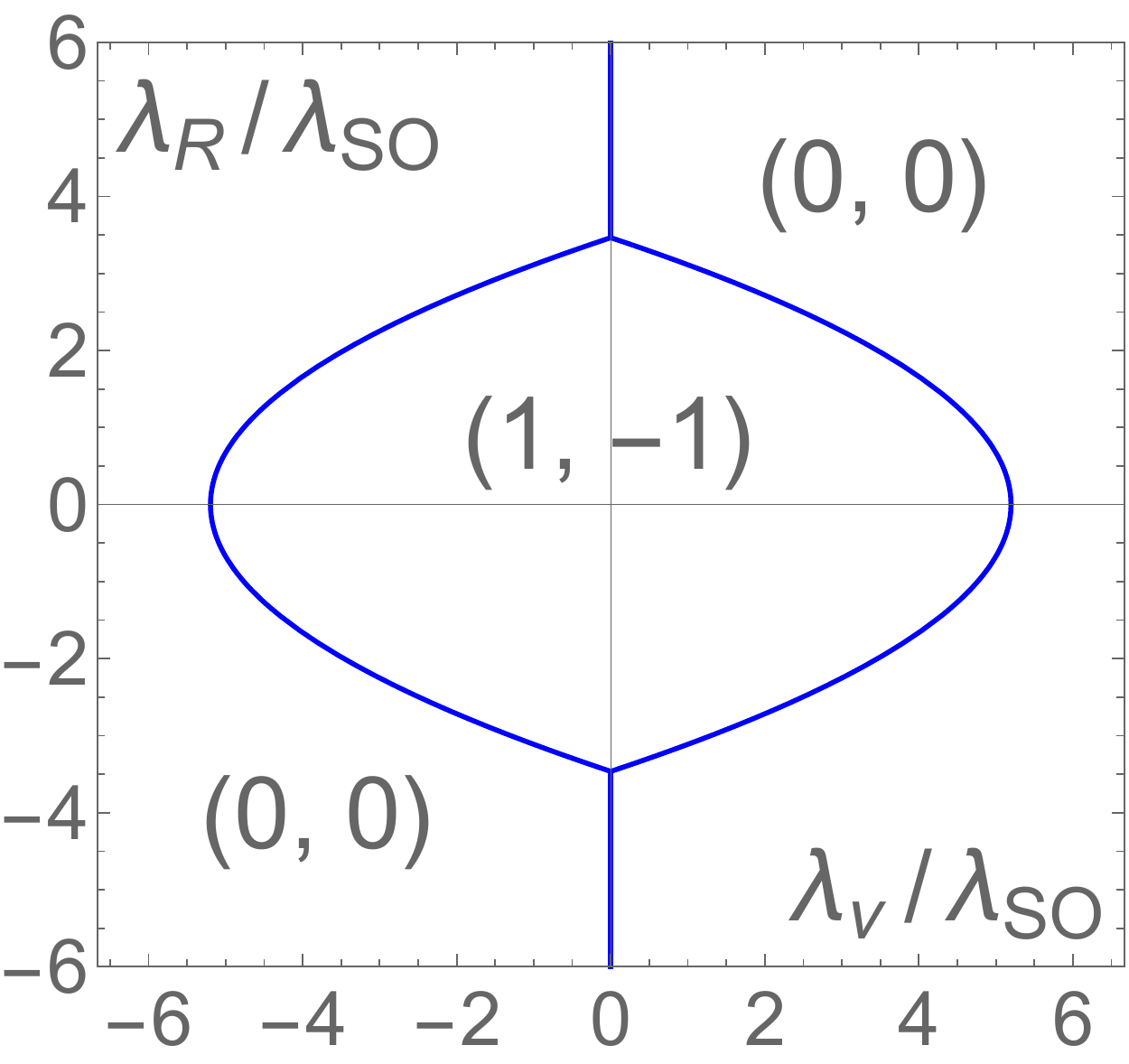}
 \end{center}
 \caption{(Color online) Phase diagram of the Kane--Mele model 
 determined by
 the entanglement spin Chern numbers (e-Ch-$\uparrow$, e-Ch-$\downarrow$)
 as a function of $\lambda_\nu$ and $\lambda_R$ for $t=1$ and
 $\lambda_{SO}= 0.06$ as in Ref.~\citen{PhysRevLett.95.146802}. 
 (e-Ch-$\uparrow$,
 e-Ch-$\downarrow$)= (1, $-1$) and 
 (0, 0) correspond to the QSH phase and the ordinary insulator phase, 
 respectively.}
 \label{fig:PDe-Ch}
\end{figure}

The Kane--Mele model has time-reversal symmetry and is
characterized by the $Z_2$ topological invariant. Namely, the $Z_2$
invariant distinguishes the QSH state and the ordinary insulating
state. Naively, the QSH phase can be understood as a state where
spin-up and -down
electrons have finite Chern numbers with opposite signs. Then,
as we have noted in the introduction, it is expected
that the
entanglement spin Chern number has an ability to detect QSH states,
since it is defined so that the focus is on either the up- or down-spin
sector. Hereafter, we use the symbol
e-Ch-$\sigma$ to represent the entanglement spin Chern number for the
case that spin-$\bar{\sigma}$ is traced out. Figure~\ref{fig:PDe-Ch} shows the
phase diagram of the Kane--Mele
model in the $\lambda_\nu$--$\lambda_R$ plane determined by the
numerically obtained entanglement spin Chern number. The QSH phase,
which appears for small $\lambda_\nu$ and $\lambda_R$, is characterized
by (e-Ch-$\uparrow$, e-Ch-$\downarrow$)=(1,$-1$), whereas the ordinary
insulating phase is characterized by (e-Ch-$\uparrow$,
e-Ch-$\downarrow$)=(0,0). The entanglement spin Chern
number changes when the energy gap closes at the K- and K'-pointis in the
Brillouin zone, and it is confirmed that the obtained phase diagram is
equivalent to the one determined with the $Z_2$
invariant. It should be emphasized that when $\lambda_R$
is finite, the spin-up and -down sectors are mixed by the Rashba effect,
and the system is no longer a mere collection of independent subsystems,
although the topological classification by the entanglement 
spin Chern number is still valid.

%\section{Kane--Mele model with ferromagnetism}
Next, we break the time-reversal symmetry by introducing the Zeeman term
\begin{equation}
 \mathrm{H}_Z=-B_0\sum_{i}c_i^{\dagger}\left(\bm{n}\cdot\bm{s}\right) c_i
\end{equation}
into the Kane--Mele model. 
A possible way to realize this situation is to
place the honeycomb lattice on a 
ferromagnetic substrate.\cite{PhysRevB.89.195303} Because of the 
time-reversal symmetry breaking, the $Z_2$ number becomes ill-defined,
while the entanglement spin Chern number is still well-defined. In
addition, the Chern number can be finite 
owing to the time-reversal
symmetry breaking. 
In this paper, we choose the vector $\bm{n}$ so that its direction is 
perpendicular to the plane.
In the following, we determine the
phase diagram of the
Kane--Mele model with the Zeeman term by making use of the Chern number
and the entanglement spin Chern number. 

\begin{figure}[htb]
 \centering
 \includegraphics[width=8.5cm]{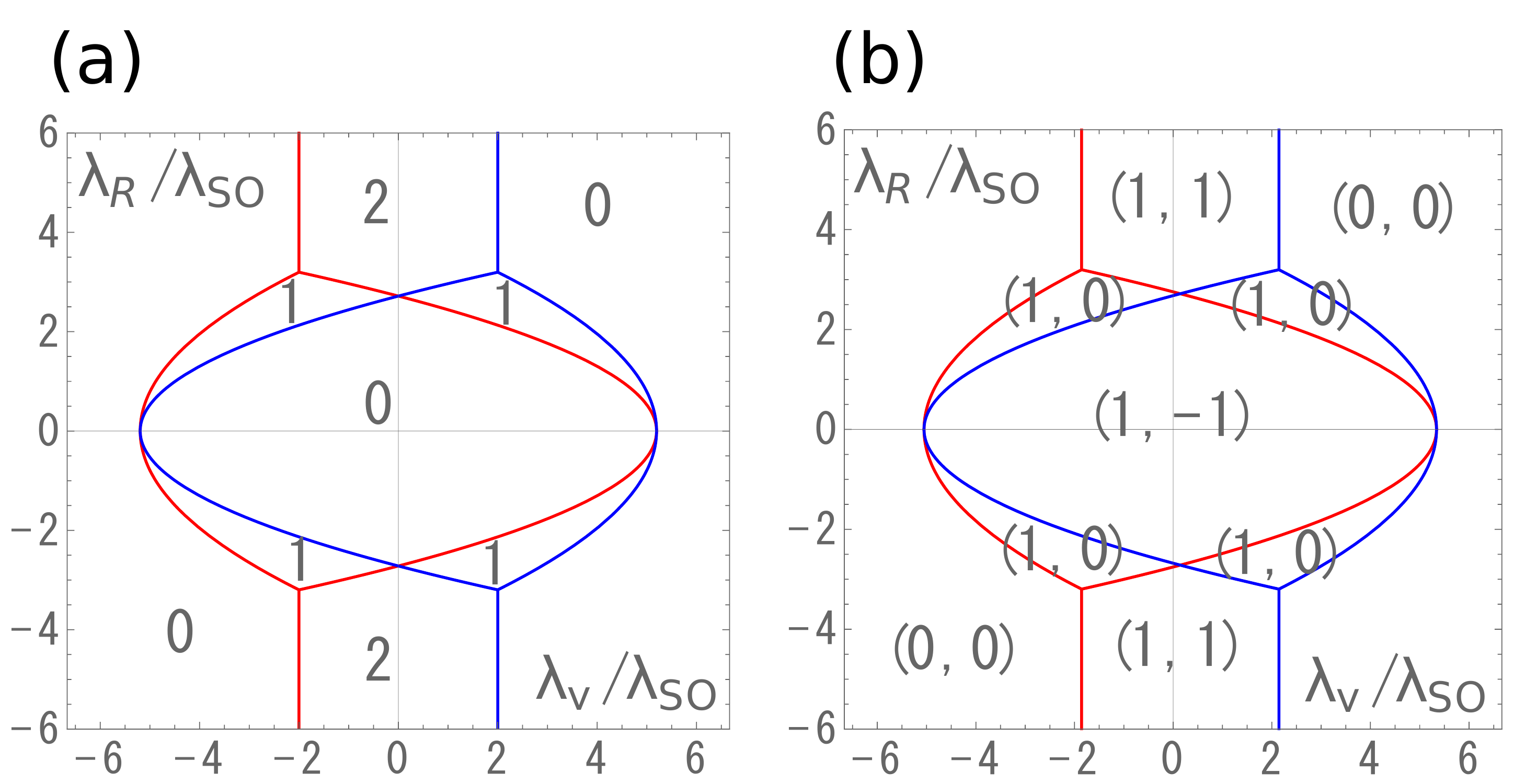}
 \caption{(Color online) Phase diagram determined by (a) Chern number
 and by (b) entanglement spin Chern number (e-Ch-$\uparrow$,
 e-Ch-$\downarrow$). In both cases
 $\lambda_{SO}/t=0.06$ and $B_0 = 2\lambda_{SO}$.
 The red (blue) line denotes the boundary of the phases 
 where the gap of the energy band closes at the K-point (K'-point).}
 \label{fig:PhaseDiagrams}
\end{figure}

Figure~\ref{fig:PhaseDiagrams} shows the phase diagrams
determined by the Chern
number and the entanglement spin Chern
number.  In Fig. 2(a), there are phases with
the Chern numbers
0, 1, and 2. In order to observe a change in the integer
topological invariant, the band gap should be closed somewhere in
the Brillouin zone. On the red (blue) lines in the phase diagram, the gap of
the energy band closes at the K-point
(K'-point). 
When the gap of the energy band closes at this point,
the gap of the entanglement spectrum with the spin partition
also closes at the same point.
Typically, the
Chern number changes by 1 (or $-1$) 
across the gap-closing line.
However, there are exceptions, namely, there are lines dividing
the phases with the Chern numbers 0 and 2, which will be
discussed later. If the Zeeman term is turned off, the Chern
number should be zero on the entire phase space,
although there are several gap-closing lines in the phase
space. The Zeeman term induces a split of the gap-closing line into 
two gap-closing lines due to the inequivalence between the 
K-point and K'-point, 
and a finite Chern number is observed in the region
surrounded by the lines. 

A basically identical phase diagram can be 
obtained by using the
entanglement spin Chern numbers. The phases with the
Chern numbers 1 and 2
correspond to the phases with (e-Ch-$\uparrow$, e-Ch-$\downarrow$)=(1,0)
and (1,1), respectively. The phase with the Chern number 0 is somewhat
special, i.e., it corresponds to (e-Ch-$\uparrow$,
e-Ch-$\downarrow$)=(0,0) \textit{and} (1,$-1$). That is,
two phases with the same Chern number are sometimes distinguished by the
entanglement Chern number. Whether the
distinction between the (0,0) and (1,$-1$) states is meaningful
even without
time-reversal symmetry is an interesting future
subject. It is worth
noting that the sum rule, a rule that the Chern number is the sum of
e-Ch-$\uparrow$ and e-Ch-$\downarrow$, holds in this case. 

%Furthermore, we set $\vec{n}$ parallel to the honycomb plane.
%In this case, the phase diagram of the entanglement Chern number is almost same as the case of the perpendicular magnetic field.
%Despite the time-reversal symmetry breaking, the Chern number is trivial in these systems.

\begin{figure}[htb]
 \centering
 \includegraphics[scale=0.25]{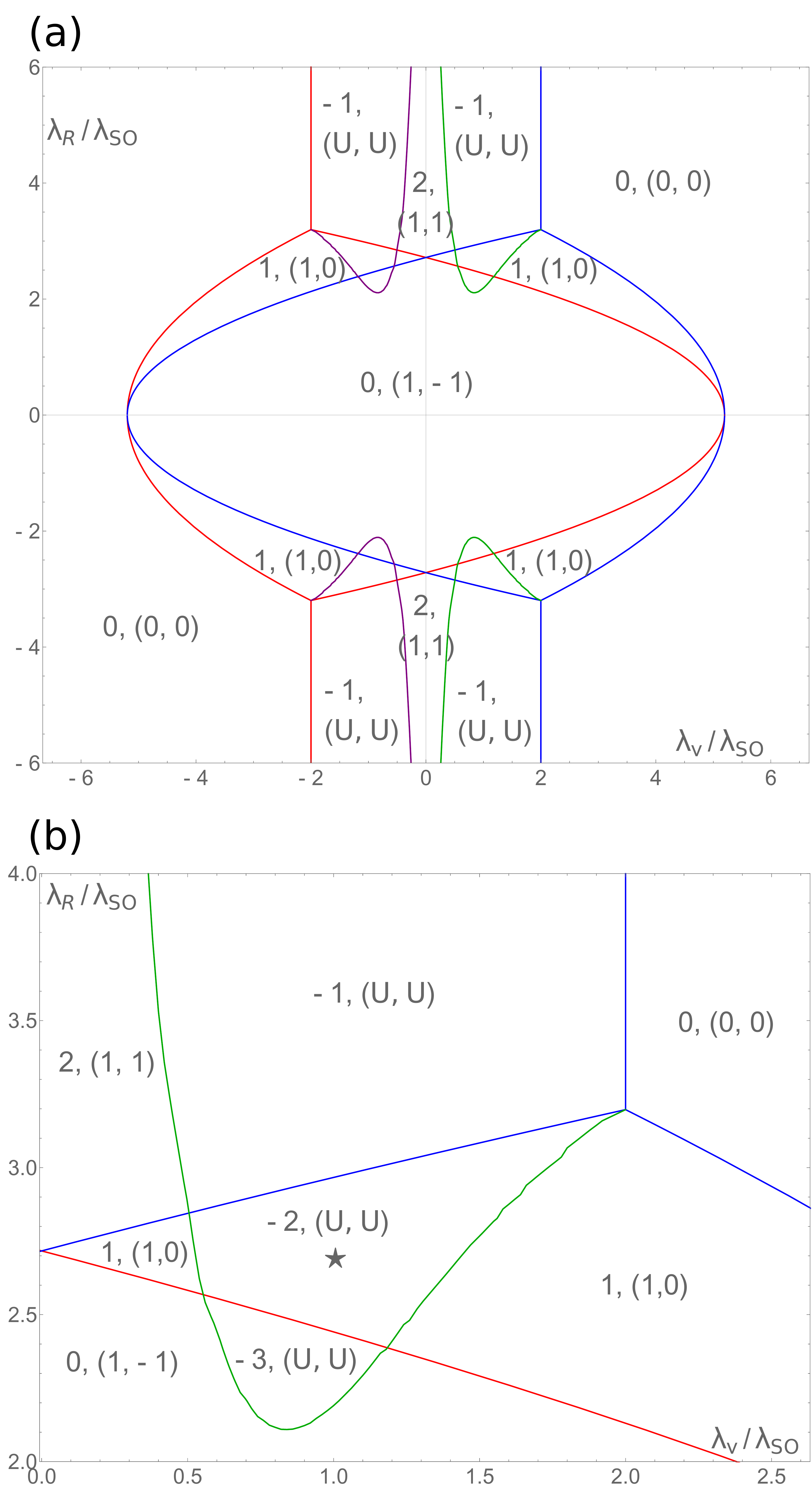}
 \caption{(Color online) 
 (a) Phase diagram for $\lambda_{SO}=0.6t$ and $B_0 = 2\lambda_{SO}$. 
 The Chern number and the entanglement spin Chern numbers 
 (e-Ch-$\uparrow$, e-Ch-$\downarrow$) are indicated in each phase.
 (b) Magnified view of the phase diagram.
 The symbol ``U'' denotes the region where the entanglement Chern number is undefined since the 
 entanglement spectrum is gapless (see Fig. \ref{fig:e-spectrum}).}
 \label{fig:PhaseDiagramLargeSO}
\end{figure}

Let us consider the case of a large spin-orbit coupling.
In this case, the energy dispersion is semimetallic
i.e., the hole and electron bands overlap in the energy
space.\cite{PhysRevB.90.165136} However, provided the
``direct gap'' is always finite over the entire Brillouin zone, the
Chern number is
well-defined, and the phase diagram determined by the Chern 
number is depicted in Fig. \ref{fig:PhaseDiagramLargeSO}.
In this case, novel phases with a negative Chern number
are revealed.
On the red (blue)
lines in the phase diagram, the
gap of the energy band closes at the K-point (K'-point), as
in the small spin-orbit coupling limit. On the other hand,
on the purple 
(green) lines, the energy band closes at a point on the $\Gamma$-K 
($\Gamma$-K'). 
On these lines, the gap of the entanglement spectrum with the spin 
partition also closes at a point where the energy band closes.
Reflecting the three fold rotational symmetry, three Dirac cones 
appear in the energy dispersion of the system on the purple and green line.
A single Dirac cone contributes a value of $\pm 1/2$ to the Chern
number; \cite{PhysRevLett.53.2449} thus,
the Chern number changes by 3
across the purple and green lines.

In the small spin-orbit coupling limit
($\lambda_{so}\ll t$), the purple (green) line
becomes much closer to the red (blue) line.
Then, the phases with a negative Chern number become invisible. 
This explains the existence of 
phase boundaries with the Chern numbers 0 and 2 in Fig. 2.
Each of them is actually a pair of boundaries where one divides the
phases with Chern numbers 2 and $-1$ 
and the other divides the phases with Chern numbers $-1$
and 0.

\begin{figure}[htb]
 \centering
 \includegraphics[scale=0.60]{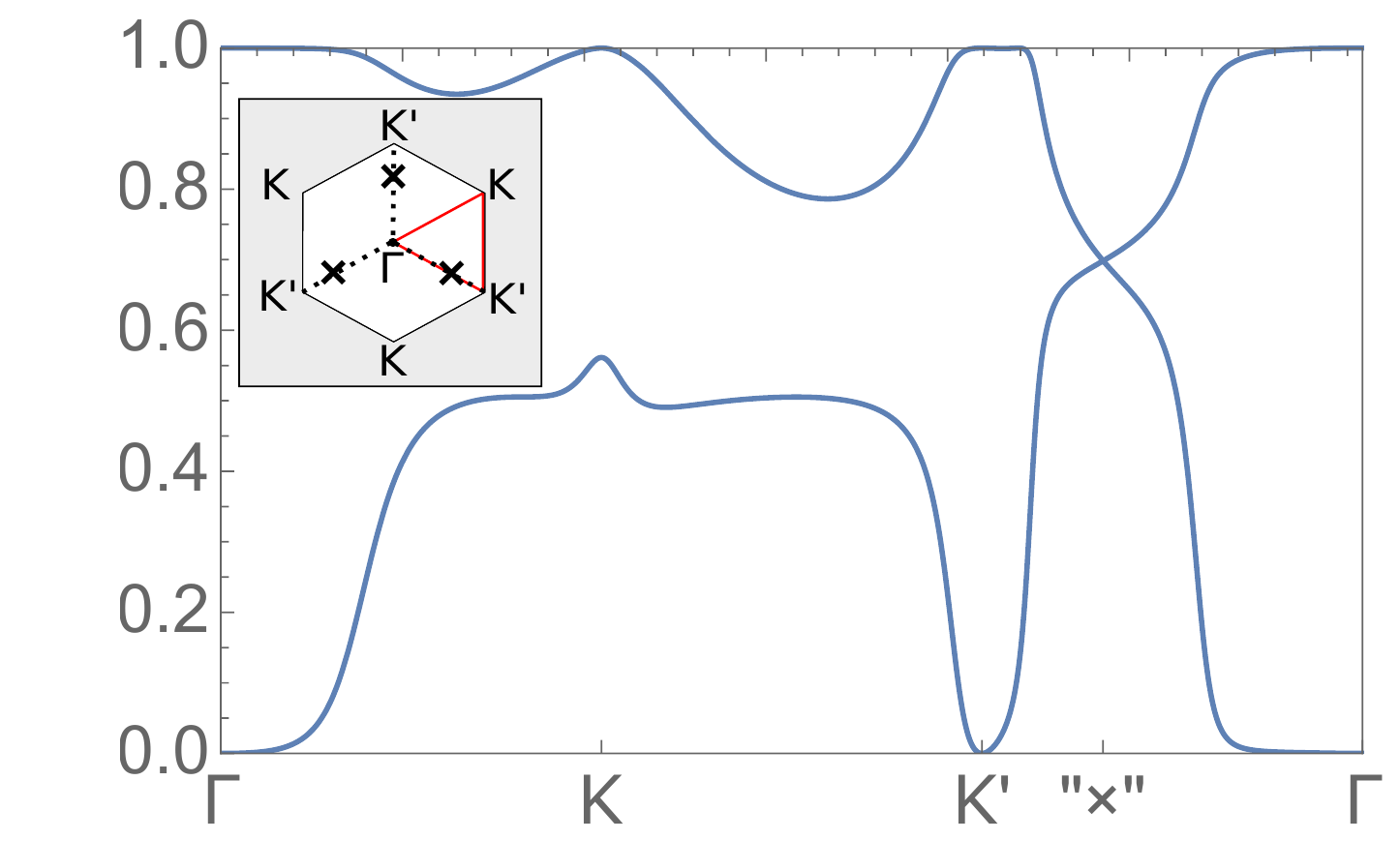}
 \caption{(Color online) Spin-up entanglement spectrum 
 with $\lambda_{SO}=0.6t$, $B_0 = 2\lambda_{SO}$, and 
 $(\lambda_R,\lambda_\nu)=(2.7\lambda_{SO},1.0\lambda_{SO})$, which
 correspond to the black star in Fig.
 \ref{fig:PhaseDiagramLargeSO}(b).
 (Inset) First Brillouin zone.
 The gap is closed at the three points indicated by ``$\times$''. }
 \label{fig:e-spectrum}
\end{figure}

In the phases with a negative Chern number, the entanglement spin Chern
number is undefined since the entanglement spectrum is gapless.
We have shown the entanglement spectrum
in Fig. \ref{fig:e-spectrum} for the gapless case
$(\lambda_R,\lambda_\nu)=(2.7\lambda_{SO},1.0\lambda_{SO})$.
In this  parameter region of negative Chern numbers, 
the spectrum of the entanglement Hamiltonian
is always gapless. This gap-closing momentum is continuously shifted and becomes 
gapped at the phase boundaries specified by the green lines in Fig. \ref{fig:PhaseDiagramLargeSO}. 
This phenomenon occurs for the system regardless of the value 
of $\lambda_{SO}$.
This implies that the spin partition is not suitable for this model
in the parameter region. 
Except in the region with a negative Chern number, there is again 
correspondence between the entanglement Chern number and the Chern
number, i.e., the sum rule holds as in the case of $\lambda_{so} \ll t$.

%\section{Summary and Outlook}
To summarize, we demonstrated that the entanglement Chern number is
useful for characterizing the QSH states in the case of 
time-reversal invariance
and the nonconservative of $s_z$. The results are consistent with the
characterization by the $Z_2$ invariant. Next we found a case in which
phases with the same Chern number result in a different
entanglement spin Chern number even in the case of broken time-reversal 
symmetry, for instance, the states with
(e-Ch-$\uparrow$, e-Ch-$\downarrow$)=(0, 0) and (1, $-1$) as in
Fig.~\ref{fig:PhaseDiagrams}.
Investigating the significance of this difference is
an important future issue. Another important finding of this paper is
the sum rule that the sum of the entanglement Chern numbers
is equal to the original Chern number. Although this sum rule is
empirical, it is ideal to have a solid theoretical explanation. The
sum rule also applies to the case without time-reversal symmetry. In
addition, when the time-reversal symmetry is broken, we
find a finite region in the phase diagram where the entanglement spin
Chern number is undefined owing to the gap closing in the entanglement
spectrum, despite the fact that the gap in the energy dispersion remains
finite. 
We should clarify whether this phenomenon is physical or is
an artifact caused by an unsuitable choice of the partition.
Generally, the entanglement Chern number depends on the partition. 
Therefore, one needs to use the most suitable partition to
obtain the topological properties of a many-body ground state. 
It reminds us of the order parameters for describing the phase transitions,
which are crucially important for choosing a suitable partition.

There are also many variants of the entanglement Chern number, i.e., we
can apply not only the spin partition but also the orbital partition,
the sublattice partition, the layer-by-layer partition, \cite{PhysRevB.93.115106}
and so on. Thus,
the concept of the entanglement Chern number introduces many types of
topological invariants. It is an interesting task to give intuitive
understanding of the known topological phases and to explore new 
phases by making use of the entanglement Chern number. 

\section*{Acknowledgments}
This work was supported in part by JSPS KAKENHI Grant Numbers
26247064 and 25400388.

%\nocite{*}
%\bibliographystyle{jpsj}
%\bibliography{17164}

\end{document}